\documentclass[aps,prab,twocolumn,superscriptaddress,nofootinbib]{revtex4-2}

\usepackage{physics}
\usepackage{amsfonts}
\usepackage{graphicx}
\usepackage{verbatim}
\usepackage{xcolor}
\usepackage{soul}
\SetSymbolFont{largesymbols}{bold}{OMX}{txex}{b}{n}
\usepackage{bm}
\newcommand{\mat}[1]{\mathsf{#1}}
\newcommand{\red}[1]{{\color{black}{#1}}}

\begin{document}


\title{Generalization of the Froissart-Stora formula to piecewise-linear spin-orbit resonance crossings}


\author{Joseph P. Devlin}
\email{jpd285@cornell.edu}
\affiliation{Department of Physics, Cornell University, Ithaca, NY 14853, USA}
\author{Georg H. Hoffstaetter}
\affiliation{Department of Physics, Cornell University, Ithaca, NY 14853, USA}
\affiliation{Brookhaven National Laboratory, Upton, NY 11973, USA}
\author{Desmond P. Barber}
\affiliation{Deutsches Elektronen-Synchrotron DESY, Notkestra\ss{}e 85, 22607 Hamburg, Germany}
\affiliation{Department of Mathematics and Statistics, University of New Mexico, Albuquerque, NM 87131, USA}


\date{May 29, 2026}

\begin{abstract}
Spin-polarized beams are important for some nuclear and high-energy physics experiments, such as those planned for the future Electron-Ion Collider (EIC). However, maintaining polarization during the acceleration of a charged-particle beam is difficult because the periodic nature of circular accelerators leads to spin-orbit resonances where the spin-precession frequency is a sum of integer multiples of the orbital frequencies. Usually, the dominant depolarization mechanisms are first-order spin-orbit resonances and the depolarization associated with crossing such a resonance can be computed using the Froissart-Stora formula. However, accelerating polarized hadron beams to high energy requires special magnet structures called Siberian snakes. When these are implemented to maintain a spin-precession frequency of one-half the revolution frequency, there will be no first-order spin-orbit resonance crossings. The dominant depolarization mechanisms are then higher-order spin-orbit resonances. The Froissart-Stora formula can be applied to higher-order resonances when the slope of the amplitude-dependent spin tune is constant. However, the slope of the amplitude-dependent spin tune often changes at the moment of resonance crossing. This work introduces a generalization of the Froissart-Stora formula which is applicable when the slope changes in this manner. The applicability of this formula is demonstrated through tracking simulations of a higher-order resonance crossing in both a toy model and the Relativistic Heavy Ion Collider (RHIC). It is additionally shown that the Froissart-Stora formula is mathematically equivalent to the Landau-Zener formula for the diabatic transition probability in two-level systems with a linearly increasing energy gap and constant coupling. This work therefore also extends the Landau-Zener formula to the case of changing slope. 
\end{abstract}


\maketitle

\section{Introduction}
The Electron-Ion Collider (EIC), which is currently being designed for construction at Brookhaven National Laboratory (BNL), will probe the internal structure of nucleons through collisions of {spin-}polarized electron beams with beams of {spin-}polarized protons, {spin-}polarized light nuclei, and unpolarized heavy nuclei~\cite{EIC}. The Hadron Storage Ring (HSR) of the EIC will accelerate not only polarized proton beams, as is already done in BNL's Relativistic Heavy Ion Collider (RHIC)~\cite{RHIC}, but also polarized beams of helions and possibly other species. The primary obstacle in maintaining polarization during acceleration of these beams will be the crossing of spin-orbit resonances.

Spin-orbit resonances occur when the spin motion, governed by the
Thomas-Bargmann-Michel-Telegdi (T-BMT) equation, is in resonance with the orbital motion so that changes in the spin motion due to the electromagnetic fields along a particle's trajectory build up coherently~\cite{Jackson,Thomas,BMT}. The spins may then no longer align with the required direction so that the beam loses some polarization. 
For hadrons in flat rings containing no special magnets for controlling the spin motion, the strength of the effect is mostly well-described in terms of the 
amplitudes of Fourier harmonics (the ``resonance strengths") describing the basic coupling of the spin motion to the orbital motion.
Methods of computing such resonance strengths are well-known, and the associated depolarization from resonance crossing can be calculated using the Froissart-Stora formula~\cite{F-S, Devlin_IPAC_24}. 

However, in rings with Siberian snakes, such as RHIC and the HSR, there are no first-order resonance crossings and one needs to consider higher-order spin-orbit resonances. This naturally leads to the goal of defining resonance strengths for such resonances in an attempt to exploit the Froissart-Stora formula for those cases as well. This  paper will show how to extend the use of the Froissart-Stora formula to higher-order resonances more generally than has been possible in the past~\cite{Higher_Order}. Additionally, the mathematical equivalence of the Froissart-Stora formula and the Landau-Zener formula is presented so that this work also provides a generalization of the Landau-Zener formula.

\section{Spin dynamics}

The normalized expectation value of the spin operator evaluated in the particle's rest frame (simply called ``the spin"), denoted by $\vec{S}$, evolves according to the T-BMT equation, which takes the form
\begin{equation}
\dv{t}\vec{S}(t)=\vec{W}(\vec{E},\vec{B},\vec{p},t)\times\vec{S}(t),
\end{equation}
where $\vec{E},\vec{B},\vec{p}$, and $t$ are the electric field, magnetic field, momentum, and time, respectively, all of which are evaluated in the lab frame~\cite{Jackson,Thomas,BMT}. The beam polarization is defined as the magnitude of the ensemble-averaged spin. In accelerator physics, where the fields are a known function of the position along the reference trajectory, the phase-space coordinate of the particle, and the system parameters, the T-BMT equation is commonly written in the form
\begin{equation}
\label{eq:T-BMT}
\begin{split}
\dv{\theta}\vec{S}(\theta)&=\vec{\Omega}\bm{(}\vec{z}(\theta),\theta;\tau\bm{)}\times\vec{S}(\theta),
\\
\vec{\Omega}\bm{(}\vec{z}(\theta),\theta;\tau\bm{)}&=\frac{L}{2\pi}\qty[\frac{hp}{vp_l}\vec{W}\bm{(}\vec{z}(\theta),\theta;\tau\bm{)}-\vec{g}\times\vec{e}_z],
\end{split}
\end{equation}
where $L$, $\vec{g}$, and $\vec{e}_z$ are the total length, curvature, and tangential unit vector of the design orbit, respectively; $p=\|\vec{p}\|$ and $v=\|\vec{v}\|$ are the magnitudes of the the particle momentum and velocity; $p_l=\vec{p}\cdot\vec{e}_z$; and $h=1+g_xx+g_yy$~\cite{Georg}. The new independent variable $\theta$ is the generalized machine azimuth, which is just a rescaling of the distance along the design orbit (usually denoted {by} $s$), and $\vec{z}(\theta)\in\mathbb{R}^6$ is the vector of phase-space coordinates at azimuth $\theta$~\cite{Vogt}. The system parameters, e.g., the reference energy and the orbital tunes, which often change slowly during the ramping process, are characterized by $\tau$. 


In this study, we assume that the orbital motion is integrable so that it can be described by three sets of action-angle variables $(\vec J, \vec \Phi)$ with tunes $\vec Q$, which in general depend on $\vec J$ (the ``amplitude-dependent tunes").  In order to study spin-orbit resonances, we also need to define a tune for the spin motion. This task is not trivial because the spins do not precess uniformly in the machine coordinate system. Nevertheless, we can find the proper tune by describing the motion with a Hamiltonian and transforming to action-angle variables {as in~\cite{Yokoya_Action_Angle}.} 
For the spin motion, in analogy with the orbital motion, the tune
is the frequency of spin precession divided by a particle's
circulation frequency on the closed orbit.
Note that, in principle, the orbital motion is influenced by the spin motion through a quasi-Stern-Gerlach effect, but since the associated 
potential energy is minuscule compared to the particle energy in the HSR, it can be neglected for our purposes~\cite{Vogt,ZfP1,ZfP2}. The orbital 
Hamiltonian and the spin Hamiltonian can therefore be handled separately, with the spin 
treated as a spectator.  

The action variable for the spin motion is the projection of the spin onto a classical quantization axis 
denoted by $\vec{n}(\vec{z},\theta;\tau)$~\cite{D-K,Yokoya_Action_Angle}.
The \emph{invariant spin field} (ISF) $\vec{n}(\vec{z},\theta;\tau)$ is a field over $(\vec z,\theta)$ of normalized solutions to the T-BMT equation which is $2\pi$-periodic in $\theta$, i.e., 
\begin{equation}
\begin{split}
\dv{\theta}\vec{n}\bm{(}\vec{z}(\theta),\theta;\tau\bm{)}&=\vec{\Omega}\bm{(}\vec{z}(\theta),\theta;\tau\bm{)}\times\vec{n}\bm{(}\vec{z}(\theta),\theta;\tau\bm{)},
\\
\vec{n}(\vec{\xi},\theta;\tau) &=  \vec{n}(\vec{\xi},\theta + 2 \pi;\tau),
\end{split}
\end{equation}
for any orbital trajectory $\vec{z}(\theta)$ and phase-space position $\vec{\xi}$. Note that the ISF is defined for \emph{constant} $\tau$, so if the system parameters were changing continuously, the ISF would be a different, pre-determined vector field at every instant. The projection 
\begin{equation}
J_S(\theta) = \vec{S}(\theta)\cdot\vec{n}\bm{(}\vec{z}(\theta),\theta;\tau(\theta)\bm{)}
\end{equation}
is called the spin action, and for constant $\tau$, it is invariant along a particle trajectory since $\vec S$ and
$\vec{n}$ then both obey the constant-$\tau$ T-BMT equation, which conserves dot products~\cite{Georg,Vogt}. The ISF is a central object for classifying spin motion~\cite{Hoffstaetter-Vogt-Barber,Quasiperiodic}.

The angle variable for the spin motion describes the orientation of a spin's
projection onto 
a plane orthogonal to the ISF. This plane is spanned by two unit vectors $\vec{u}_1(\vec{z},\theta;\tau)$ and $\vec{u}_2(\vec{z},\theta;\tau)$, defined at constant $\tau$ such that $(\vec{u}_1,\vec{u}_2,\vec{n})$ is a right-handed, orthonormal coordinate system for all $(\vec{z},\theta)$. As with $\vec{n}$, the vector fields $\vec{u}_1$ and
$\vec{u}_2$ are $2\pi$-periodic in $\theta$, and they are chosen so that the projection precesses at a uniform rate $\nu$, which
is the number of spin precessions around $\vec n$ for one turn around the accelerator, and is therefore referred to as the \emph{amplitude-dependent spin tune} (ADST)~\cite{Quasiperiodic}. The ADST $\nu=\nu(\vec{J};\tau)$ can depend on the orbital amplitudes and the system parameters. Any solution to the T-BMT equation which begins orthogonal to the ISF will remain orthogonal to the ISF and precess $\nu$ oscillations around $\vec{n}$ per turn. Therefore, for $\vec{u}_1$ and $\vec{u}_2$ to be periodic, they must obey the equation
\begin{equation}
\begin{split}
\dv{\theta}\vec{u}_i\bm{(}\vec{z}(\theta),\theta;\tau\bm{)}=&\qty[\vec{\Omega}\bm{(}\vec{z}(\theta),\theta;\tau\bm{)}-\nu\,\vec{n}\bm{(}\vec{z}(\theta),\theta;\tau\bm{)}]
\\
&\times \vec{u}_i\bm{(}\vec{z}(\theta),\theta;\tau\bm{)}.
\end{split}
\end{equation}
The coordinate system $(\vec{u}_1,\vec{u}_2,\vec{n})$ is called the \emph{uniform invariant frame field} (u-IFF) because spins precess uniformly in it and it is periodic in $\theta$~\cite{Quasiperiodic}. 

The spin-precession vector can be separated into two parts as 
\begin{equation}
\vec{\Omega}(\vec{z},\theta;\tau)=\vec{\Omega}_0(\theta;\tau)+\vec{\omega}(\vec{z},\theta;\tau),
\end{equation}
where $\vec{\Omega}_0$ is the spin-precession vector on the closed orbit and $\vec{\omega}$ includes fields resulting from particle oscillations about the closed orbit. In a circular accelerator, the lattice is periodic so that 
$\vec{\Omega}_0(\theta;\tau)=\vec{\Omega}_0(\theta+2\pi;\tau)$. 

On the closed orbit, $\vec n$  is denoted by $\vec n_0$. It depends only on $\theta$ and $\tau$, and it is $2\pi$-periodic in $\theta$. It is therefore easily computed as the rotation axis of the one-turn {3~$\times$~3} spin-transfer matrix on the closed orbit. In a flat ring, $\vec n_0$ is vertical everywhere. In the 
presence of solenoids or horizontal magnetic
fields on the closed orbit (e.g., due to misalignments),  $\vec n_0$ will be tilted from the vertical. The closed-orbit counterparts to $\vec{u}_1(\vec{z},\theta;\tau)$ and $\vec{u}_2(\vec{z},\theta;\tau)$ {can be chosen to coincide with} $\vec{m}(\theta;\tau)$ and $\vec{l}(\theta;\tau)$, the $2\pi$-periodic unit vectors of the SLIM formalism, which in this case are {defined} so that spins precess around $\vec{n}_0$ at a uniform rate in the $(\vec{m},\vec{l},\vec{n}_0)$ coordinate system~\cite[Section 2.6.6]{Handbook3}. This rate is the closed-orbit spin tune $\nu_0$, and in a flat ring, $\nu_0=G\gamma$, where $G$ is the anomalous gyromagnetic ratio and $\gamma$ is the Lorentz factor~\cite{Georg}. For general rings, the fractional part of $\nu_0$ can be extracted from the complex eigenvalues $\exp(\pm i2\pi\nu_0)$ of the one-turn spin-transfer matrix on the closed orbit. {With our choice of $\vec{u}_1$ and $\vec{u}_2$, the ADST reduces to $\nu_0$ as $\vec{z}$ approaches the closed orbit.}

Whereas the calculation of $\vec n_0$ and $\nu_0$ is straightforward, the calculation
of $\vec{n}$ and $\nu$ for general phase-space trajectories requires sophisticated methods such as stroboscopic averaging, SODOM-2, normal form analysis, adiabatic anti-damping, or Fourier analysis~\cite{Strobe_Origin,SODOM2,Hoffstaetter-Vogt-Barber,Quasiperiodic,Hamwi_NAPAC_25}.
However, as we shall see later, there is a powerful model which allows for an analytical calculation of $\vec{n}$.
This is the so-called {\emph{single-resonance model}} (SRM)~\cite{Courant_Ruth, Mane, Quasiperiodic} and, as we shall see, it is the central tool in this paper.

Spins that are not parallel to the ISF rotate around it with frequency $\nu$. When the precession vector ${\vec \omega}(\vec z, \theta;\tau)$ changes with the system parameters $\tau$, the spin motion generally does not change dramatically if ${\vec \omega}$ only has Fourier terms with frequency $f\neq\nu$, because then the products of spin components and $\vec\omega$ components average to zero. Dramatic changes are only expected when $\vec\omega$ has components that also oscillate with frequency $\nu$. It is therefore of interest to determine which frequencies comprise the Fourier spectrum of $\vec\omega$. 

The function $\vec\omega(\vec{z},\theta;\tau)$ is $2\pi$-periodic in $\theta$ and in the orbital angles $\vec \Phi$ and can therefore be written as a Fourier series with exponents $k_0\theta + \vec k\cdot \vec \Phi$. Hence, it contains a Fourier component with frequency $\nu$ at the resonance condition
\begin{equation}
\label{eq:intrinsic}
\nu(\vec J;\tau)=k_0+\vec{k}\cdot\vec{Q}(\vec{J};\tau),
\end{equation}
where $(k_0,\vec{k})\in\mathbb{Z}^4$. The order of the resonance is defined as $\|\vec{k}\|_1$. 
Near such resonances,  $\vec n$ can be strongly tilted from $\vec n_0$ by an amount that usually increases with the relevant orbital amplitudes. In nominally flat rings with small misalignments, $\nu$ is usually close to $\nu_0$ so that it increases 
with the beam energy and can pass through the resonance condition of Eq.~\eqref{eq:intrinsic}
during acceleration. First-order resonances ($\|\vec{k}\|_1=1$) are usually called intrinsic resonances. An analogous concept exists for motion on the closed orbit because, in the presence of misalignments, $\vec n_0$ can be  
strongly tilted from the design direction when $\nu_0$ is near an integer.
This phenomenon is usually called an imperfection resonance.

When a system parameter changes, such as the reference energy $E_0$ during acceleration, the spin action is an adiabatic invariant along particle trajectories, even when spin-orbit resonances are crossed~\cite{Adiabatic_Invariance}. Informally, this means that the variation of $J_S$ can be made arbitrarily small by changing the system parameters sufficiently slowly. 

Although $J_S$ is an adiabatic invariant, its value may change when polarized beams are accelerated to their storage energy and resonances are crossed at some finite rate. It is of interest to quantify the degree of invariance because the loss of spin action usually describes irreversible polarization loss. However, the converse is not true: polarization loss does not imply spin-action reduction because when the ISF varies widely over phase space, the polarization will be small even when every particle has $J_S \approx 1$. The spread of the ISF is quantified by the maximum time-averaged polarization~\cite{Georg}
\begin{equation}
P_\mathrm{lim}(\vec{J},\theta;\tau) = \norm{\frac{1}{(2\pi)^3}\int_{[0,2\pi]^3}\vec{n}(\vec{J},\vec{\Phi},\theta;\tau)\dd[3]{\vec{\Phi}}}.
\end{equation}
{Near spin-orbit resonances, the ISF tends to spread out over phase space, resulting in a small $P_\mathrm{lim}$.}

The long-term loss of spin action encountered when crossing an isolated spin-orbit resonance at a constant rate is well-approximated by the Froissart-Stora formula~\cite{F-S}. \red{In low-to-medium-energy rings, the first-order resonances are isolated well enough that the Froissart-Stora formula predicts the level of depolarization very well, as demonstrated by its application to various machines in the decades following its publication~\cite{FS1,Courant_Ruth,SYLee,FS2}.}

In rings with Siberian snakes which are used to force $\nu_0=1/2$ independently of energy, all first-order resonances are avoided because the tunes are kept away from half-integers for orbital stability~\cite[Fig.~9]{Montague}. However, even if $\nu_0$ does not depend on the system parameters $\tau$, $\nu(\vec{J};\tau)$ can still change with $\tau$. Then, even though $\nu$ usually remains near 1/2, it can cross higher-order resonance conditions $(\|\vec{k}\|_1>1)$.

\begin{figure}[h]
\centering
\includegraphics[scale=0.65]{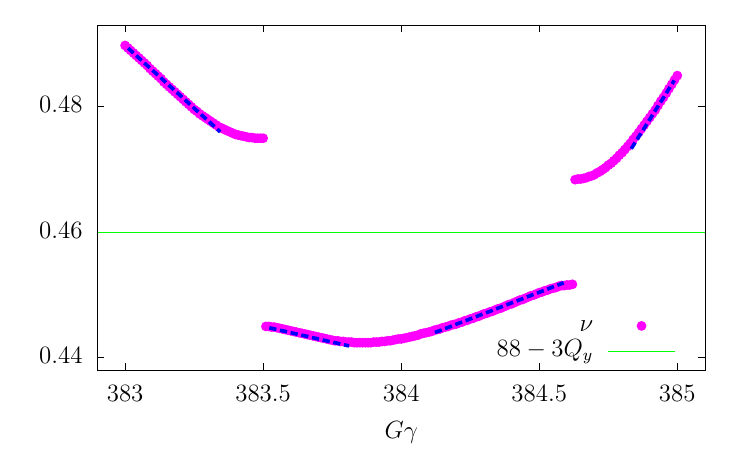}
\caption{\label{fig:Res_ex}A pair of third-order spin-orbit resonances across which the slope of the ADST $\nu$ changes drastically. The spin-orbit resonance crossings are characterized by discontinuities in $\nu$ and the dashed lines indicate the approximate slopes.}
\end{figure}

It has been shown that the Froissart-Stora formula can describe the loss of spin action due to crossing of these higher-order resonances if 
$\nu(E_0)$
is linear in the vicinity of the resonance crossing~\cite{Higher_Order}. However, the functional form of $\nu(E_0)$ is a property of the lattice and the particle amplitude, and it is more common that $\nu(E_0)$ is approximately piecewise linear, with different slopes before and after the crossing, as shown in Fig.~\ref{fig:Res_ex}. When $E_0(\theta)$ is ramped linearly, as is usual when the beam is accelerated to its storage energy, $\nu(\theta)$ will then also be piecewise linear. Therefore, in this paper, we will derive an extension to the Froissart-Stora formula which applies when the slope changes at the moment of resonance crossing, and we will show how this formula can be used to estimate the loss of spin action due to higher-order resonance crossings for which a decent estimate was previously not possible.

\section{The resonance spectrum}

\subsection{Resonance strengths}
As mentioned in the previous section, in a circular accelerator, the lattice periodicity implies that the components of $\vec{\omega}$ are functions of $\theta$ whose spectra are subsets of~\cite[Section 4]{Quasiperiodic}
\begin{equation}
\label{eq:frequency_set}
\mathcal{F}= \qty{k_0+\vec{k}\cdot\vec{Q}:(k_0,\vec{k})\in\mathbb{Z}^4}.
\end{equation}

Away from resonances, the polarization usually remains near $\vec{n}_0$. Then, the part of $\vec{\omega}\bm{(}\vec{z}(\theta),\theta\bm{)}$ which drives depolarization is orthogonal to $\vec{n}_0$. It can be represented by the complex quantity
\begin{equation}
\label{eq:little_omega}
\omega\bm{(}\vec{z}(\theta),\theta\bm{)} = \vec{\omega}\bm{(}\vec{z}(\theta),\theta\bm{)}\cdot\qty[\vec{m}(\theta)+i\vec{l}(\theta)].
\end{equation}
The individual Fourier components $\tilde\epsilon_\kappa$ comprising the spectrum of the quasiperiodic function in Eq.~\eqref{eq:little_omega} can be recovered from
\begin{equation}
\label{eq:Resonance_Strength_Definition}
\tilde{\epsilon}_\kappa=\lim_{\Theta\to\infty}\frac{1}{\Theta}\int_0^\Theta\omega\bm{(}\vec{z}(\theta),\theta\bm{)}e^{-i\kappa\theta}\dd{\theta},
\end{equation}
where $\kappa\in\mathcal{F}$~\cite{Quasiperiodic}. 
Each complex Fourier component has an amplitude $\epsilon_\kappa$ and a phase $\phi_\kappa$. 
Then, with $\mathcal K = \|\vec{k}\|_1$, 
we call $\epsilon_\kappa$ the ${\mathcal K}^{\rm {th}}$-{\emph{order resonance strength}} with frequency $\kappa$ and phase $\phi_\kappa$~\cite{Georg}. 

The primary spin-orbit resonances studied in storage rings {without Siberian snakes} are first-order resonances driven by linear vertical betatron motion, where the most important components of $\vec{\omega}\bm{(}\vec{z}(\theta),\theta\bm{)}$ result from the horizontal magnetic fields in the quadrupoles. We will therefore now focus on the case of $\|\vec{k}\|_1=1$.

\subsection{The single-resonance model (SRM)}
When $\nu_0 \approx \kappa$ for some $\kappa\in\mathcal{F}$ and the remaining elements of $\mathcal{F}$ are sufficiently far from $\kappa$, it is reasonable to make the approximation 
\begin{equation}
\omega\bm{(}\vec{z}(\theta),\theta\bm{)} \approx \epsilon_\kappa e^{i(\kappa\theta+\phi_\kappa)},
\end{equation}
as the other Fourier components will contribute negligibly to the spin motion. This approximation is the basis of the SRM.

When only first-order resonances resulting from vertical betatron motion are being studied, the SRM is commonly defined by the spin-precession vector
\begin{equation}
\label{eq:SRM_Definition}
\vec{\omega}\bm{(}\vec{z}(\theta),\theta\bm{)}=\epsilon\qty[\cos(\kappa\theta+\phi)\vec{m}(\theta)+\sin(\kappa\theta+\phi)\vec{l}(\theta)],
\end{equation}
where $\kappa = k_0\pm Q$ with $k_0\in\mathbb{Z}$ and $Q$ is the vertical betatron tune. Note that the vertical betatron motion could be replaced by the horizontal betatron motion or the synchrotron motion, although their influence is usually less significant {when $\vec{n}_0$ is nearly vertical}.

This model has a constant precession rate $\nu_0$ around $\vec{n}_0$ on the closed orbit with respect to the chosen 
$(\vec{m},\vec{l})$. The additional precession due to the orbital motion is represented by $\vec\omega$, which rotates with frequency $\kappa$ in the plane orthogonal to $\vec{n}_0$. 

The u-IFF and the ADST in the SRM can be derived analytically 
by transforming from the $(\vec{m}$, $\vec{l}, \vec{n}_0)$ basis into a basis rotated around $\vec{n}_0$ by an angle $-(\kappa\theta+\phi)$~\cite{Quasiperiodic, Georg, Mane, Devlin_IPAC_24}. In the new basis, the precession vector is
\begin{equation}
\vec{\Omega}_R=\begin{pmatrix} \epsilon \\ 0 \\ \nu_0-\kappa \end{pmatrix}.
\end{equation}
{A spin vector parallel to $\vec{\Omega}_R$ is constant in this frame and satisfies the T-BMT equation. Hence, it has the required periodicity of the ISF. The ISF can be obtained in any other basis by inverting the prior rotation.} In addition, since the ISF is only defined up to a sign, we will add the sign factor $\operatorname{sgn}(\delta)$ to the ISF so that $\vec{n}\cdot\vec{n}_0>0$ always. One can then construct $\vec{u}_1$ and $\vec{u}_2$ as
\begin{equation}
\begin{split}
\vec{u}_2&=\frac{\vec{n}_0\times\vec{n}}{\|\vec{n}_0\times\vec{n}\|},
\\
\vec{u}_1&=\vec{u}_2\times\vec{n}.
\end{split}
\end{equation}
In general, such a construction produces an orthonormal basis $(\vec{u}_1,\vec{u}_2,\vec{n})$, but it may not produce a u-IFF, i.e., a basis in which spins precess uniformly. Incidentally, for the SRM, this procedure does produce a u-IFF, but the precession frequency in this frame is $\operatorname{sgn}(\delta)\sqrt{\epsilon^2+\delta^2}$, which does not reduce to $\nu_0$ for $\epsilon = 0$. Therefore, we will choose to rotate $\vec{u}_1$ and $\vec{u}_2$ around $\vec{n}$ by an angle $-(\kappa\theta+\phi)$, which creates another u-IFF. The precession frequency in the resulting u-IFF does reduce to $\nu_0$ for $\epsilon=0$.

In total, our choice for the u-IFF expressed in the 
($\vec{m}$, $\vec{l}$, $\vec{n}_0$) basis is
\begin{equation}
\label{eq:SRM_Solution}
\begin{split}
\vec{u}_1\bm{(}\vec{z}(\theta),\theta\bm{)}&=\frac{\operatorname{sgn}(\delta)}{\Lambda}\begin{pmatrix} |\delta|\cos^2\Phi+\Lambda\sin^2\Phi \\ (|\delta|-\Lambda)\sin\Phi\cos\Phi \\ -\epsilon\cos\Phi \end{pmatrix},
\\
\vec{u}_2\bm{(}\vec{z}(\theta),\theta\bm{)}&=\frac{\operatorname{sgn}(\delta)}{\Lambda}\begin{pmatrix} (|\delta|-\Lambda)\sin\Phi\cos\Phi \\ \Lambda\cos^2\Phi + |\delta|\sin^2\Phi \\ -\epsilon\sin\Phi \end{pmatrix},
\\
\vec{n}\bm{(}\vec{z}(\theta),\theta\bm{)}&=\frac{\operatorname{sgn}(\delta)}{\Lambda}\begin{pmatrix} \epsilon\cos\Phi \\ \epsilon\sin\Phi \\ \delta \end{pmatrix},
\\
\nu(\vec{J})&=\operatorname{sgn}(\delta)\Lambda+\kappa,
\end{split}
\end{equation}
where $\Phi=\kappa\theta+\phi$, $\delta = \nu_0-\kappa$, and $\Lambda = \sqrt{\epsilon^2+\delta^2}$. Note that, for $\epsilon \neq 0$, $\nu$ is not continuous as a function of $\nu_0$, but rather jumps by $2\epsilon$ at $\nu_0=\kappa$. Here, $\nu_0$ should be regarded simply as a parameter of the model. Using Eq.~\eqref{eq:SRM_Solution}, one can show that the maximum time-averaged polarization for this model takes the simple form
\begin{equation}
P_\mathrm{lim}(\vec{J})=\frac{|\delta|}{\Lambda}.
\end{equation}
As expected, this expression vanishes for $\nu_0=\kappa$ and increases as one moves away from the resonance.

Although it suffices in most cases, this focus of the SRM on first-order resonances with a single orbital degree of freedom is not actually simpler than the general case. If we allow $\kappa$ to be $k_0+\vec{k}\cdot\vec{Q}$ for any $\|\vec{k}\|_1$, then Eq.~\eqref{eq:SRM_Solution} still holds.

\subsection{The Froissart-Stora formula}
\label{sec:derivation}
The change of $J_S$ due to a resonance crossing can be computed explicitly for the SRM~\cite{F-S}. The mathematical manipulation is made simpler by the use of complex spinors $\Psi$ rather than real three-vectors. Equation~\eqref{eq:T-BMT} is equivalent to the equations
\begin{equation}
\label{eq:spinors}
\begin{split}
\vec{S}(\theta)&=\Psi^\dag(\theta)\vec{\sigma}\Psi(\theta),
\\
i\Psi'(\theta)&=\frac{1}{2}\qty[\vec{\Omega}\bm{(}\vec{z}(\theta),\theta\bm{)}\cdot\vec{\sigma}]\Psi(\theta),
\end{split}
\end{equation}
where $\vec{\sigma}$ is the vector of Pauli matrices and $'$ indicates the derivative {of} a function with respect to its argument. To ensure the normalization of $\vec{S}$, $\Psi$ must be normalized in the sense of the 2-norm on $\mathbb{C}^2$~\cite{Courant_Ruth}.

Beginning with Eqs.~\eqref{eq:SRM_Definition} and \eqref{eq:spinors}, and then calculating in the $(\vec{m},\vec{l},\vec{n}_0)$ basis, the equation of motion for the spinor $\Psi$ in the SRM is
\begin{equation}
\label{eq:SRM_EOM}
\Psi'(\theta) = -\frac{i}{2}\begin{pmatrix} \nu_0 & \epsilon e^{-i(\kappa\theta+\phi)} \\ \epsilon e^{i(\kappa\theta+\phi)} & -\nu_0 \end{pmatrix}\Psi(\theta).
\end{equation}
The Froissart-Stora formula treats the case of linearly ramped $\nu_0$ with a resonance crossing at $\theta=0$, i.e., ${\nu_0=\kappa+\alpha\theta}$. Hence,
\begin{equation}
\Psi'(\theta) = -\frac{i}{2}\begin{pmatrix} \kappa+\alpha\theta & \epsilon e^{-i(\kappa\theta+\phi)} \\ \epsilon e^{i(\kappa\theta+\phi)} & -(\kappa+\alpha\theta) \end{pmatrix}\Psi(\theta).
\end{equation}
We now transform to the interaction picture with~\cite{Mane_Two_Resonances}
\begin{equation}
\label{eq:FS_IP}
\Psi_I(\theta)=\exp\bm{\bigg(}\frac{i}{2}\qty(\kappa\theta+\frac{1}{2}\alpha\theta^2)\sigma_3\bm{\bigg)}\,\Psi(\theta).
\end{equation}
In the $(\vec{m},\vec{l},\vec{n}_0)$ basis, we write $\vec{S}=s_1\vec{m}+s_2\vec{l}+s_3\vec{n}_0$. It is important to note that $\Psi^\dag\sigma_3\Psi=\Psi_I^\dag\sigma_3\Psi_I$, so we can simply use $\Psi_I$ when calculating $s_3$.
The resulting equation of motion is
\begin{equation}
\Psi_I'(\theta)=-\frac{i}{2}\begin{pmatrix} 0 & \epsilon e^{i(\alpha\theta^2/2-\phi)} \\ \epsilon e^{-i(\alpha\theta^2/2-\phi)} & 0 \end{pmatrix}\Psi_I(\theta).
\end{equation}
Denoting the spinor components by $\Psi_I=(\Psi_+,\Psi_-)^\mathrm{T}$, we have the {set of} coupled equations
\begin{equation}
\label{eq:Coupled_Hermite}
\Psi'_\pm(\theta)=-\frac{i}{2}\epsilon\exp\bm{\bigg(}\pm i\qty(\frac{1}{2}\alpha\theta^2-\phi)\bm{\bigg)}\,\Psi_\mp(\theta).
\end{equation}
These equations can be combined into a second-order, uncoupled equation for either component. We choose to eliminate $\Psi_-$:
\begin{equation}
\label{eq:Before_Transformation_Hermite}
\Psi''_+(\theta)- i\alpha\theta\Psi'_+(\theta)+\frac{\epsilon^2}{4}\Psi_+(\theta)=0.
\end{equation}

With the substitution $x=\sqrt{i\alpha/2}\theta$, Eq.~\eqref{eq:Before_Transformation_Hermite} becomes the Hermite equation:
\begin{equation}
\begin{split}
&\Psi_+(\theta)=\widetilde{\Psi}(x),
\\
&\widetilde{\Psi}''(x)-2x\widetilde{\Psi}'(x)-\frac{i\epsilon^2}{2\alpha}\widetilde{\Psi}(x)=0.
\end{split}
\end{equation}
Two linearly independent solutions are
\begin{equation}
\begin{split}
\widetilde{F}(x)&={}_1F_1\qty(\frac{i\epsilon^2}{8\alpha};\frac{1}{2};x^2),
\\
\widetilde{G}(x)&=x\,{}_1F_1\qty(\frac{i\epsilon^2}{8\alpha}+\frac{1}{2};\frac{3}{2};x^2),
\end{split}
\end{equation}
where ${}_1F_1(a;b;z)$ is the confluent hypergeometric function of the first kind~\cite{Abramowitz_Stegun}. Defining $F(\theta)=\widetilde{F}(x)$ and the same for $G$, the general solution of Eq.~\eqref{eq:Before_Transformation_Hermite} is
\begin{equation}
\label{eq:General_Solution}
\Psi_+(\theta)=C_1F(\theta)+C_2G(\theta).
\end{equation}
The derivatives of $F$ and $G$ can be calculated using the relation
\begin{equation}
\dv{z}\,{}_1F_1(a;b;z)=\frac{a}{b}\,{}_1F_1(a+1;b+1;z).
\end{equation}

Let
\begin{equation}
\mat{M}(\theta) = \begin{pmatrix} F(\theta) & G(\theta) \\ F'(\theta) & G'(\theta) \end{pmatrix},
\end{equation}
whereby
\begin{equation}
\begin{pmatrix} \Psi_+(\theta) \\ \Psi_+'(\theta) \end{pmatrix} = \mat{M}(\theta)\begin{pmatrix} C_1 \\ C_2 \end{pmatrix}.
\end{equation}
The initial condition used in the Froissart-Stora formula is $\Psi_+(-\theta_0)=1$ and $\Psi'_+(-\theta_0)=0$ with $\theta_0>0$, where the initial condition on $\Psi_+'$ follows necessarily from Eq.~\eqref{eq:Coupled_Hermite} and the normalization of $\Psi$. Then,
\begin{equation}
\begin{split}
\begin{pmatrix} C_1 \\ C_2 \end{pmatrix} &=
\mat{M}(-\theta_0)^{-1} \begin{pmatrix} 1 \\ 0 \end{pmatrix},
\\
\begin{pmatrix} \Psi_+(\theta_0) \\ \Psi_+'(\theta_0) \end{pmatrix} &= \mat{M}(\theta_0)\mat{M}(-\theta_0)^{-1} \begin{pmatrix} 1 \\ 0 \end{pmatrix}.
\end{split}
\end{equation}
The asymptotics of $F$ and $G$ follow from the fact that for large $z$ on the positive real axis, we have~\cite[Eq.~(5.3.51)]{Morse_Feshbach}
\begin{equation}
{}_1F_1(a;b;iz) \sim \frac{\Gamma(b)}{\Gamma(a)}z^{a-b}i^{a-b}e^{iz}+\frac{\Gamma(b)}{\Gamma(b-a)}z^{-a}i^a.
\end{equation}
Using this result, one can show that
\begin{equation}
\lim_{\theta_0\to\infty}|\Psi_+(\theta_0)|^2 = \exp(-\frac{\pi\epsilon^2}{2\alpha}).
\end{equation}
This yields the Froissart-Stora formula:
\begin{equation}
\lim_{\theta_0\to\infty}s_3(\theta_0)=2\exp(-\frac{\pi\epsilon^2}{2\alpha})-1.
\end{equation}

The Froissart-Stora formula rigorously describes the change of $\vec{S}\cdot\vec{n}_0$ when $\nu_0$ is ramped linearly. However, as $\vec{n}\to\vec{n}_0$ for $|\theta| \to \infty$ and $\nu \approx \nu_0$ away from the spin-tune jump, the Froissart-Stora formula also approximately describes the change of $J_S$ when $\nu$ is ramped linearly (apart from the jump by $2\epsilon$ across $\nu=\kappa$). 

We now show the connection between the Froissart-Stora formula and the Landau-Zener formula, as pointed out in~\cite{Turrin}. Note that the latter was derived nearly three decades before the former~\cite{LZ_Landau,LZ_Zener,LZ_Stuckelberg,LZ_Majorana}. In terms of the notation of~\cite[Section 28.5]{Zwiebach}, and with $\hbar=1$, we consider the two-level system governed by the Schr\"odinger equation
\begin{equation}
\label{eq:LZ_Hamiltonian}
i\psi'(t)=\begin{pmatrix}\tilde\alpha t/2 & H_{12} \\ H_{12}^* & -\tilde\alpha t/2 \end{pmatrix}\psi(t).
\end{equation}
{The system described by this equation has} two states with a linearly increasing energy gap $\tilde{\alpha}t$ and constant coupling $H_{12}$. With the wavefunction components denoted by $\psi=(\psi_+,\psi_-)^\mathrm{T}$, the Landau-Zener formula states that, given $\psi_+(-t_0)=1$, 
\begin{equation}
\label{eq:LZ_Formula}
\lim_{t_0\to\infty}|\psi_+(t_0)|^2=\exp(-\frac{2\pi|H_{12}|^2}{\tilde\alpha}),
\end{equation}
which is reminiscent of the Froissart-Stora formula.

The connection to the Froissart-Stora formula is made explicit by the transformation
\begin{equation}
\Psi_\mathrm{LZ}(\theta)=\exp(\frac{i}{2}\kappa\theta\sigma_3)\Psi(\theta).
\end{equation}
We again denote the components by $\Psi_\mathrm{LZ}=(\Psi_+,\Psi_-)^\mathrm{T}$. This transformation retains the important property $\Psi^\dag\sigma_3\Psi=\Psi_\mathrm{LZ}^\dag\sigma_3\Psi_\mathrm{LZ}$.
The resulting equation of motion is
\begin{equation}
\label{eq:LZ_Motion}
\Psi_\mathrm{LZ}'(\theta)=-\frac{i}{2}\begin{pmatrix} \alpha\theta & \epsilon e^{-i\phi} \\ \epsilon e^{i\phi} & -\alpha\theta \end{pmatrix}\Psi_\mathrm{LZ}(\theta).
\end{equation}
This equation is in the form of Eq.~\eqref{eq:LZ_Hamiltonian} with $\tilde\alpha=\alpha$ and $|H_{12}|=\epsilon/2$. Therefore, by Eq.~\eqref{eq:LZ_Formula},
\begin{equation}
\lim_{\theta_0\to\infty}|\Psi_+(\theta_0)|^2=\exp(-\frac{\pi\epsilon^2}{2\alpha}).
\end{equation}
This result matches that of Froissart and Stora.

\section{Higher-Order Spin-Orbit Resonances}
\label{sec:higher_order}
\subsection{Applicability of the Froissart-Stora formula}
Estimates with the Froissart-Stora formula show that with typical first-order resonance strengths and acceleration rates, the polarization loss would be unacceptable during acceleration to high energy in  colliders like RHIC~\cite{Georg}. Observations already confirm such behavior at lower energies.
As indicated earlier, polarization loss  can be significantly suppressed  by installing magnet systems called Siberian snakes in combinations which constrain $\nu_0$ to 1/2 independently of $G\gamma$ so that first-order resonances can be avoided~\cite{Georg}.
A Siberian snake is a magnet system which rotates spins by $180^\circ$  around some axis while having negligible effect on the closed orbit in the rest of the ring. RHIC utilizes pairs of snakes with positions and axes chosen so that $\nu_0=1/2$ while also ensuring that $\vec{n}_0$ is vertical in the arcs.
For reasonable orbital amplitudes, the ADST then remains in a small band around 1/2. However, although first-order resonances can be avoided entirely by choosing
orbital tunes far from 1/2, it is known that there can still be significant depolarization due to higher-order resonances.

Spin-orbit resonances of arbitrary order are usually characterized by a discontinuity in $\nu$, which {mirrors the behavior of the ADST} in the SRM. It is therefore tempting to define the higher-order resonance strength as half the size of the spin-tune jump and use this in the Froissart-Stora formula to estimate the change in $J_S$ when $\nu$ is ramped linearly. However, not all resonances are contained in the spectrum of $\vec{\omega}$, as demonstrated by the fact that higher-order resonances can appear even when $\vec{\omega}$ is a linear function of $\vec{z}$~\cite{Vogt,Georg,ManeA,ManeB}. The spin motion near such resonances therefore cannot be similar to the SRM. For these cases, a more general discussion of spin motion near an isolated resonance is required.

We can mathematically motivate the application of the Froissart-Stora formula to higher-order resonances with the following argument. When some system parameter $\tau$ changes, the evolution of the u-IFF at each $(\vec z, \theta)$ can only be a rotation around a vector $\vec{\eta}(\vec{z},\theta;\tau)${, i.e.,}
\begin{equation}
\pdv{\tau}\vec{n}(\vec{z},\theta;\tau)=\vec{\eta}(\vec{z},\theta;\tau)\cross\vec{n}(\vec{z},\theta;\tau),
\end{equation}
with identical equations for $\vec{u}_1$ and $\vec{u}_2$~\cite{Georg}. For the linear variation $\tau=\alpha\theta$, the equations of spin motion take the form~\cite{Higher_Order}
\begin{equation}
\dv{\theta}\begin{pmatrix} s_1 \\ s_2 \\ J_S \end{pmatrix} = \begin{pmatrix} \alpha(\eta_3s_2-\eta_2J_S)-\nu(\vec{J};\tau)s_2 \\ \alpha(\eta_1J_S-\eta_3s_1)+\nu(\vec{J};\tau)s_1 \\ \alpha(\eta_2s_1-\eta_1s_2) \end{pmatrix},
\end{equation}
where $\vec{S}=s_1\vec{u}_1+s_2\vec{u}_2+J_S\vec{n}$ and $\vec{\eta}=\eta_1\vec{u}_1+\eta_2\vec{u}_2+\eta_3\vec{n}$. In terms of the complex quantities $\hat{s} = s_1+is_2$ and $\eta = \eta_1+i\eta_2$, the first two equations can be combined into the single equation
\begin{equation}
\label{eq:general_s_hat}
\hat{s}'=i\qty[\nu(\vec{J};\tau)-\alpha\eta_3]\hat{s}+i\alpha\eta J_S.
\end{equation}
Within the SRM, when $\tau$ is chosen to be $\delta = \nu_0-\kappa$, $\vec{\eta}$ can be computed explicitly for each $\delta \neq 0$. The resulting equation of motion for $\hat s$ is~\cite{Higher_Order}
\begin{equation}
\label{eq:SRM_s_hat}
\hat{s}'=i\qty[\operatorname{sgn}(\delta)\Lambda+\kappa]\hat{s}+\operatorname{sgn}(\delta)\alpha\frac{\epsilon}{\Lambda^2}e^{i(\kappa\theta+\phi)}J_S.
\end{equation}
This equation is not valid at $\nu_0=\kappa$ since the u-IFF is discontinuous there by choice.

When there is a spin-tune jump, we define $\kappa^*$ as the center frequency of the spin-tune jump and the higher-order resonance strength $\epsilon_{\kappa^*}$ as half the size of the spin-tune jump. To put Eq.~\eqref{eq:general_s_hat} into the form of Eq.~\eqref{eq:SRM_s_hat}, we define $\Lambda^* = |\nu-\kappa^*|$, whereby $\nu=\operatorname{sgn}(\nu-\kappa^*)\Lambda^*+\kappa^*$. Additionally, when $\nu \approx \kappa^*$ and the frequencies in a spectral representation of $\eta$ are well-separated, we can approximate $\eta$ by the single Fourier component $\eta_{\kappa^*}e^{i(\kappa^*\theta+\phi)}$. Equation~\eqref{eq:general_s_hat} can then be written as
\begin{equation}
\label{eq:general_higher_order}
\hat{s}'=i\qty[\operatorname{sgn}(\nu-\kappa^*)\Lambda^*+\kappa^*-\alpha\eta_3]\hat{s}+i\alpha\eta_{\kappa^*}e^{i(\kappa^*\theta+\phi)} J_S.
\end{equation}
If $\alpha\eta_3\hat{s}$ is sufficiently small, e.g., for slow ramping, then Eq.~\eqref{eq:general_higher_order} has approximately the same form as Eq.~\eqref{eq:SRM_s_hat} if $\eta_{\kappa^*}=-i\operatorname{sgn}(\nu-\kappa^*)\epsilon_{\kappa^*}/{\Lambda^*}^2$ in the vicinity of the resonance. There is reason to believe that $\eta_{\kappa^*}$ takes this form, because it is known from the SRM that this form will produce the spin-tune jump by $2\epsilon_{\kappa^*}$ across the resonance frequency $\kappa^*$. When these arguments are valid, the Froissart-Stora formula gives the change in $J_S$ due to crossing the resonance.

Another way of motivating the use of the SRM here can be found 
in~\cite[Section 4.8]{Vogt}, where a heuristic analysis based on spin
normal forms is invoked. The foregoing discussion essentially argues that even when the spin motion near an isolated resonance does not behave like the SRM, one can transform to a different coordinate system where the motion more closely resembles the SRM. This idea is related to the concept of a one-resonance normal form and a detailed treatment of normal forms (spin, one-resonance, and more) is contained in~\cite{Etienne}.

It has been shown that the assumptions made in these arguments can be met well enough that the Froissart-Stora formula can be successfully applied to higher-order resonance crossings~\cite{Higher_Order,Georg,Vogt}.

\subsection{The Froissart-Stora formula with unequal slopes}
One assumption made in our discussion of Eq.~\eqref{eq:general_higher_order} is that $\nu=\kappa + \alpha\theta$ in the vicinity of the resonance crossing. However, this assumption has proven to be very restrictive. \red{As pointed out in Ref.~\cite{Chao_Matrix}, when $\nu(\theta)$ is piecewise linear, the tools of Sec.~\ref{sec:derivation} are still applicable. We consider the form}
\begin{equation}
\label{eq:piecewise_nu}
\nu(\theta) = \begin{cases}\kappa+\alpha_1\theta, & \theta<0 \\ \kappa+\alpha_2\theta, & \theta \geq 0, \end{cases}
\end{equation}
away from the jump by $2\epsilon$ across $\nu=\kappa$. This functional dependence can be achieved in the SRM by varying $\nu_0(\theta)$ in the same manner.

It will now be shown that one can construct an analogue to the Froissart-Stora formula for this functional dependence~\cite{Devlin_Thesis,Devlin_NAPAC_25}.  The {upper} component of the spinor $\Psi_+$ has the general solution given by Eq.~\eqref{eq:General_Solution} for constant $\alpha$. As $\alpha(\theta)$ is piecewise constant, we will construct two different solutions for constant $\alpha$ and patch them together at $\theta=0$. To simplify the notation, we include $\alpha$ as a parameter, e.g., $F(\theta) \to F(\theta;\alpha)$. We also define constants of integration such that
\begin{equation}
\Psi_+(\theta) = \begin{cases} C_1F(\theta;\alpha_1)+C_2G(\theta;\alpha_1), & \theta <0 \\ D_1F(\theta;\alpha_2)+D_2G(\theta;\alpha_2), & \theta \geq 0. \end{cases}
\end{equation}
For $\theta<0$, we know that
\begin{equation}
\begin{pmatrix} \Psi_+(\theta) \\ \Psi'_+(\theta) \end{pmatrix}=\mat{M}(\theta;\alpha_1)\begin{pmatrix} C_1 \\ C_2 \end{pmatrix}.
\end{equation}
Let the initial condition be $\Psi_+(-\theta_0)=1$ and $\Psi'_+(-\theta_0)=0$ with $\theta_0>0$, as before. For $\theta<0$, we then have
\begin{equation}
\begin{pmatrix} \Psi_+(\theta) \\ \Psi'_+(\theta) \end{pmatrix} = \mat{M}(\theta;\alpha_1)
\mat{M}(-\theta_0;\alpha_1)^{-1} \begin{pmatrix} 1 \\ 0 \end{pmatrix}.
\end{equation}
We take $\Psi_+(\theta)$ to be continuously differentiable at $\theta=0$ by ansatz. Then,
\begin{equation}
\mat{M}(0;\alpha_1)
\mat{M}(-\theta_0;\alpha_1)^{-1} \begin{pmatrix} 1 \\ 0 \end{pmatrix} = \mat{M}(0;\alpha_2) \begin{pmatrix} D_1 \\ D_2 \end{pmatrix}.
\end{equation}
Therefore,
\begin{widetext}
\begin{equation}
\begin{pmatrix} \Psi_+(\theta_0) \\ \Psi_+'(\theta_0) \end{pmatrix} = \mat{M}(\theta_0;\alpha_2)\mat{M}(0;\alpha_2)^{-1}\mat{M}(0;\alpha_1)\mat{M}(-\theta_0;\alpha_1)^{-1}\begin{pmatrix} 1 \\ 0 \end{pmatrix}.
\end{equation}
Using \textsc{Mathematica}, we find that~\cite{Mathematica}
\begin{equation}
\label{eq:Two_Slope_FS}
\begin{split}
\lim_{\theta_0\to\infty}|\Psi_+(\theta_0)|^2 &= \exp\bm{\bigg(}-\frac{\pi\epsilon^2}{8}\qty(\frac{1}{\alpha_1}+\frac{1}{\alpha_2})\bm{\bigg)}\qty|\frac{\Gamma\qty(\frac{1}{2}+\frac{i\epsilon^2}{8\alpha_2})}{\Gamma\qty(\frac{1}{2}+\frac{i\epsilon^2}{8\alpha_1})}\cosh(\frac{\pi\epsilon^2}{8\alpha_2})-\sqrt{\frac{\alpha_2}{\alpha_1}}\frac{\Gamma\qty(1+\frac{i\epsilon^2}{8\alpha_2})}{\Gamma\qty(1+\frac{i\epsilon^2}{8\alpha_1})}\sinh(\frac{\pi\epsilon^2}{8\alpha_2})|^2.
\end{split}
\end{equation}
\end{widetext}
The asymptotic value of $s_3$ can be recovered from 
\begin{equation}
\lim_{\theta_0\to\infty}s_3(\theta_0)=2\lim_{\theta_0\to\infty}|\Psi_+(\theta_0)|^2-1.
\end{equation}
For $\alpha_1=\alpha_2$, this result reduces to the Froissart-Stora formula. By the analogy of Eqs.~\eqref{eq:LZ_Hamiltonian}~and~\eqref{eq:LZ_Motion}, one can also use Eq.~\eqref{eq:Two_Slope_FS} to construct a Landau-Zener formula for piecewise-linear energy gaps.

\begin{figure}[h]
\centering
\includegraphics[scale=0.65]{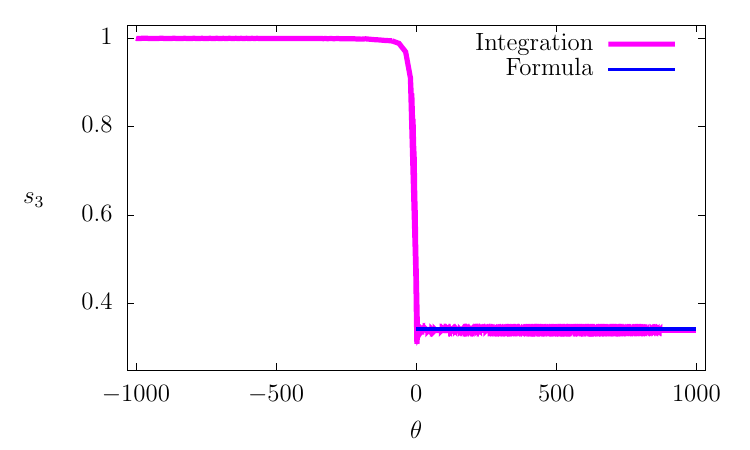}
\caption{\label{fig:Two_Slope_FS}Comparison of numerical integration of the SRM with piecewise-linear $\nu_0$ and the asymptotic value of $s_3$ predicted by Eq.~\eqref{eq:Two_Slope_FS}. The parameters are $\alpha_1=0.01$, $\alpha_2=1$, $\epsilon=0.1$, $Q=0.7$, and $\kappa=Q$.}
\end{figure}

We will first compare Eq.~\eqref{eq:Two_Slope_FS} to the result of numerical integration for a resonance crossing with vastly different slopes. We placed the SRM spin-precession vector into the T-BMT equation and varied $\nu_0$ as {in} Eq.~\eqref{eq:piecewise_nu} with $\alpha_1=0.01$, $\alpha_2=1$, $\epsilon=0.1$, $Q=0.7$, and $\kappa=Q$. We integrated from $\theta=-1000$ to $\theta=1000$ with the initial condition $\vec{S}=\vec{e}_y$. The result is shown in Fig.~\ref{fig:Two_Slope_FS} and, as expected, Eq.~\eqref{eq:Two_Slope_FS} describes the asymptotic $s_3=\vec{S}\cdot\vec{n}_0$ very well.

\begin{figure}[h]
\centering
\includegraphics[scale=0.65]{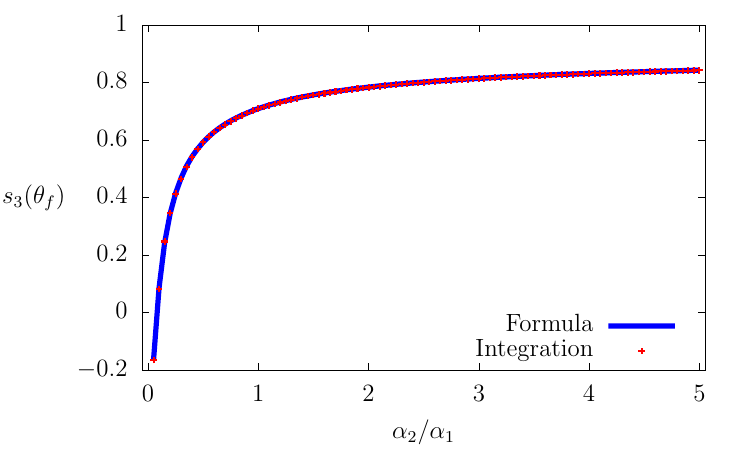}
\caption{\label{fig:Comparison}\red{Comparison of numerical integration of the SRM with piecewise-linear $\nu_0$ and the asymptotic spin projection $s_3(\theta_f)$ predicted by Eq.~\eqref{eq:Two_Slope_FS} for different slope ratios $\alpha_2/\alpha_1$. The fixed parameters are $\alpha_1=0.1$, $\epsilon=0.1$, $Q=0.7$, and $\kappa=Q$.}}
\end{figure}

\red{As an additional test of Eq.~\eqref{eq:Two_Slope_FS}, we use the same parameters except now $\alpha_1=0.1$ and $\alpha_2$ is allowed to vary. The value of $s_3$ at $\theta_f=1000$ is shown in Fig.~\ref{fig:Comparison} as a function of the ratio $\alpha_2/\alpha_1$. As before, there is very good agreement between numerical integration and Eq.~\eqref{eq:Two_Slope_FS}.}

\section{Comparison with Tracking Simulations}
\subsection{The double-resonance model with two Siberian snakes}
Attempts to create a toy model for spin motion in the presence of Siberian snakes have included the addition of a pair of Siberian snakes to the SRM. However, for properly chosen orbital tunes, the ADST does not differ meaningfully from 1/2 in this model~\cite{Mane_Exact_Solutions}. Therefore, the SRM with two Siberian snakes is not an effective model of higher-order resonance crossings. One solution is to add more Siberian snakes~\cite{Vogt}. However, there are higher-order resonances in rings such as RHIC, which only have two Siberian snakes. An alternative solution is to add a second resonance~\cite{DRM}.

\begin{figure*}
\centering
\includegraphics[scale=0.65]{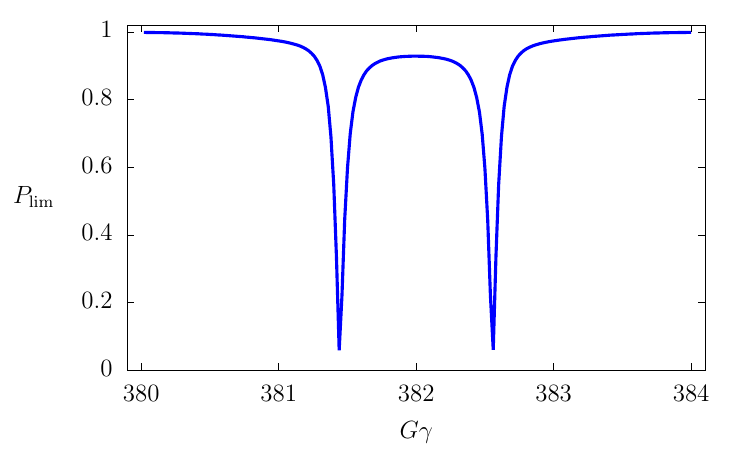}
\includegraphics[scale=0.65]{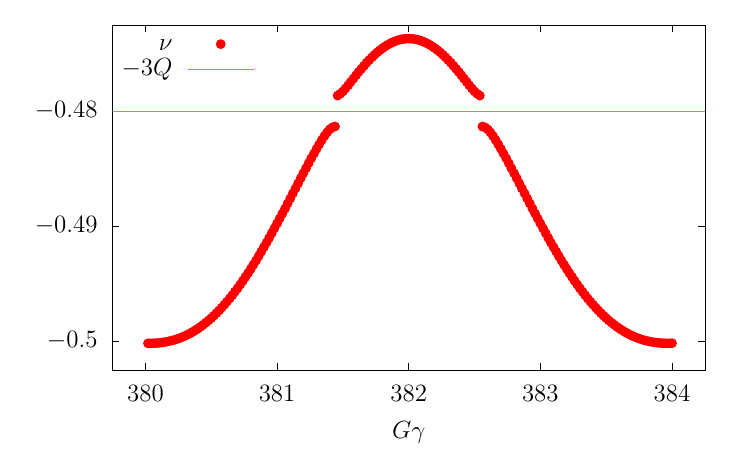}
\caption{\label{fig:DRM_FS_ADST}A pair of third-order resonances in the DRM with two Siberian snakes and parameters $Q=0.16$, $\kappa_1=382+Q$, $\kappa_2=482-Q$, $\epsilon_1=\epsilon_2=0.15$, and $\Delta\phi = \pi/4$. The resonances are characterized by a discontinuity in $\nu$ across the resonance line $-3Q$ and corresponding dips in $P_\mathrm{lim}$. Left: The maximum time-averaged polarization, computed by stroboscopic averaging. Right: The ADST, computed by averaging the rotation angle around an ISF from stroboscopic averaging~\cite{Vogt}.}
\end{figure*}

The \emph{double-resonance model} (DRM) with two  Siberian snakes and one-degree-of-freedom orbital motion is defined by the spin-precession vector
\begin{equation}
\begin{split}
\vec{\Omega}_0(\theta)&=G\gamma\vec{e}_y\nonumber
\\
&\phantom{=}+\pi\sum_{j\in\mathbb{Z}}\qty[\delta(\theta-2j\pi)\vec{e}_x+\delta\bm{(}\theta-(2j+1)\pi\bm{)}\vec{e}_z],
\\
\vec{\omega}\bm{(}\vec{z}(\theta),\theta\bm{)}&=\epsilon_1\qty[\cos(\kappa_1\theta)\vec{e}_z+\sin(\kappa_1\theta)\vec{e}_x]
\\
&\phantom{=}+\epsilon_2\qty[\cos(\kappa_2\theta+\Delta\phi)\vec{e}_z+\sin(\kappa_2\theta+\Delta\phi)\vec{e}_x],
\end{split}
\end{equation}
where $(\vec{e}_x,\vec{e}_y,\vec{e}_z)$ are radial, vertical, and longitudinal unit vectors, respectively. 
In this coordinate system, we write $\vec{S}=S_x\vec{e}_x+S_y\vec{e}_y+S_z\vec{e}_z.$ When $\Delta\phi \neq 0$, higher-order resonances can occur even in this simple model.
With this choice of snakes, $\vec n_0$ is vertically up in one half of the ring and vertically down in the 
other half.

We will now investigate whether the assumptions leading to Eq.~\eqref{eq:Two_Slope_FS} can be sufficiently satisfied for application to a higher-order resonance crossing in this model. The main concern comes from the fact that, as the ADST oscillates around 1/2 while the energy is ramped, higher-order resonances always come in pairs. For this study, the parameters were chosen to be $Q=0.16$, $\kappa_1=382+Q$, $\kappa_2=482-Q$, $\epsilon_1=\epsilon_2=0.15$, and $\Delta\phi = \pi/4$. Particle tracking and stroboscopic averaging were performed using \textsc{Bmad}~\cite{Bmad}. For this set of parameters, the higher-order resonances are centered about $G\gamma = 382$, and $P_\text{lim} \approx 0.93$ in the center. It is therefore expected that the resonances do not strongly influence the spin motion at $G\gamma=382$. The ADST and $P_{\mathrm{lim}}$
are depicted in Fig.~\ref{fig:DRM_FS_ADST}.

\begin{figure}[h]
\centering
\includegraphics[scale=0.65]{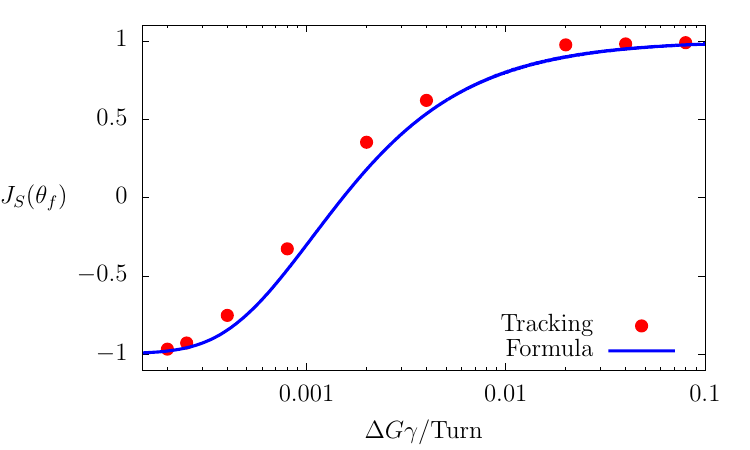}
\includegraphics[scale=0.65]{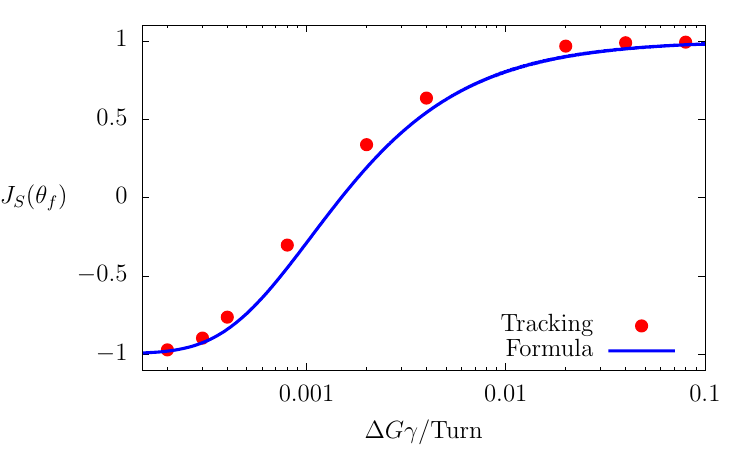}
\caption{\label{fig:DRM_FS}Comparison of the final spin action $J_S(\theta_f)$ from tracking through a higher-order resonance in the DRM with two Siberian snakes and the asymptotic spin action predicted by Eq.~\eqref{eq:Two_Slope_FS}. Top: Spin action after crossing the left resonance in Fig.~\ref{fig:DRM_FS_ADST}. Bottom: Spin action after crossing the right resonance in Fig.~\ref{fig:DRM_FS_ADST}.}
\end{figure}

We begin by tracking through the resonance on the left. To most nearly match the assumptions leading to Eq.~\eqref{eq:Two_Slope_FS}, we start tracking at $G\gamma(\theta_i)=378$ with $J_S(\theta_i)=1$. The resonances have very little influence at $\theta_i$ and the ISF is almost exactly vertical. Hence, we simply use $\vec{S}(\theta_i)=\vec{e}_y$. We stop tracking at $G\gamma(\theta_f)=382$, and we use Eq.~\eqref{eq:Two_Slope_FS} to approximate $J_S(\theta_f)$ with $\vec{n}(\theta_f)$ obtained by stroboscopic averaging~\cite{Strobe_Origin,Strobe}. The higher-order resonance strength is obtained by halving the spin-tune jump, and the slope parameter $\alpha_1$ was obtained using the formula
\begin{equation}
\alpha_1=\frac{1}{2\pi}\dv{\nu}{(G\gamma)}\eval_\text{pre-jump}\qty(\frac{\Delta G\gamma}{\mathrm{Turn}}),
\end{equation}
where $\Delta G\gamma/\mathrm{Turn}$ is the change in $G\gamma$ per turn. To compute $\alpha_2$, the pre-jump slope was replaced with the post-jump slope. The result is shown in Fig.~\ref{fig:DRM_FS}. The qualitative agreement is excellent, although the tracking and the formula do not agree exactly. This disagreement is not surprising, as the assumptions leading to Eq.~\eqref{eq:Two_Slope_FS} cannot be exactly met. Firstly, there are two resonances influencing the spin motion, although our derivation assumed that there was only one. Secondly, we tracked through a finite range surrounding the resonance, and Eq.~\eqref{eq:Two_Slope_FS} is only fully correct in the limit of tracking from infinitely far before the resonance to infinitely far after the resonance.

\begin{figure*}
\centering
\includegraphics[scale=0.65]{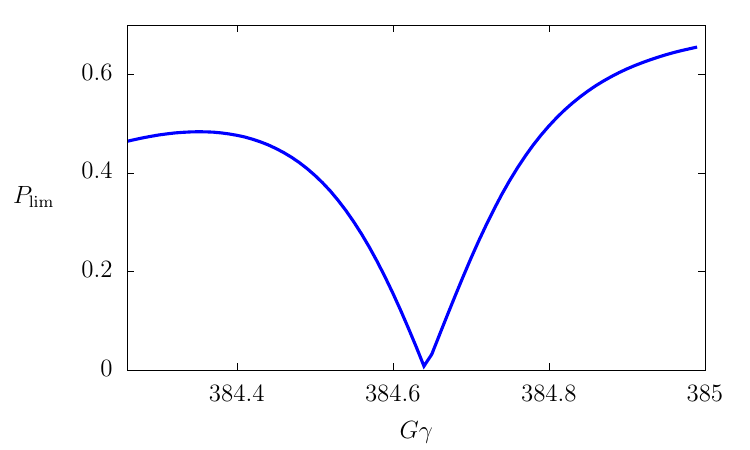}
\includegraphics[scale=0.65]{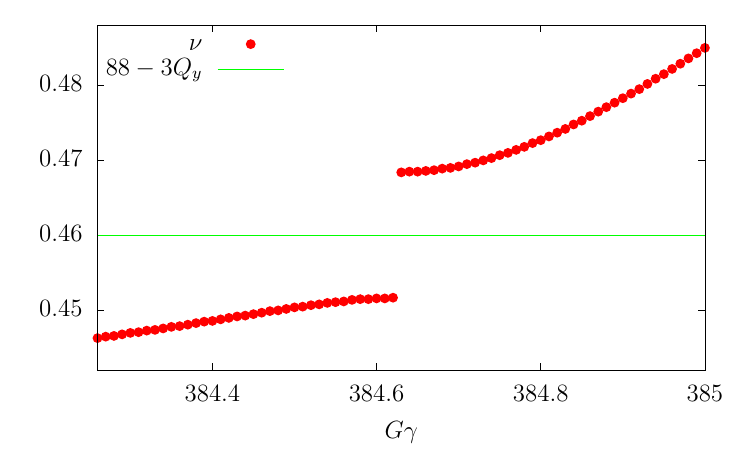}
\caption{\label{fig:RHIC_FS_ADST}Properties of spin motion for particles on the ellipse with normalized action $J_y \approx 32 \, \text{mm mrad}$ in RHIC. A third-order resonance is characterized by a discontinuity in $\nu$ across the resonance line $88-3Q_y$ and a corresponding dip in $P_\mathrm{lim}$. Left: The maximum time-averaged polarization, computed by stroboscopic averaging. Right: The ADST, computed by Fourier analysis of the turn-by-turn spin vector~\cite{Quasiperiodic}.}
\end{figure*}

We will now track through the resonance on the right. We start tracking at $G\gamma(\theta_i)=382$ with $J_S(\theta_i)=1$ and $\vec{n}(\theta_i)$ computed by stroboscopic averaging. We stop tracking at some $\theta_f$ where the influence of the resonance is almost completely absent so that $J_S(\theta_f)=S_y(\theta_f)$ to a very good approximation. We again use Eq.~\eqref{eq:Two_Slope_FS} to approximate $J_S(\theta_f)$. The result is is nearly identical to the result of tracking through the other resonance.

\subsection{The Relativistic Heavy Ion Collider}
Tracking simulations through the current model of the HSR reveal many higher-order resonances. Therefore, it would be extremely useful if Eq.~\eqref{eq:Two_Slope_FS} could be used to estimate the polarization lost when crossing higher-order resonances in a real accelerator. We will thus check the applicability of Eq.~\eqref{eq:Two_Slope_FS} to a higher-order resonance crossing in RHIC. The model of RHIC used in this work has no closed-orbit distortions or magnet errors, but third-order resonances are still visible on large-amplitude tori. This observation matches the fact that only odd-order resonances are expected for a midplane-symmetric ring~\cite{Vogt}. It is expected that a more realistic model of RHIC (with misalignments, finite-length snakes, etc.) will have higher-order resonances on a larger set of tori, as is the case with the current model of the HSR.

Our model of RHIC uses the storage optics with the $\beta$-squeeze, and the betatron tunes are $(Q_x=28.19,Q_y=29.18)$. 
The Siberian snakes are approximated by an instantaneous spin rotation with no orbital change. All tracking uses a single proton in the blue ring. There is neither horizontal betatron motion nor synchrotron motion and the normalized vertical {(Courant-Snyder)} action is 32 mm mrad. For reference, the normalized RMS vertical emittance of a polarized proton beam in RHIC is usually on the order of 1 mm mrad~\cite{RHIC_Emittance}. All tracking and stroboscopic averaging calculations were performed using \textsc{Bmad}~\cite{Bmad}.

\begin{figure}[h]
\centering
\includegraphics[scale=0.65]{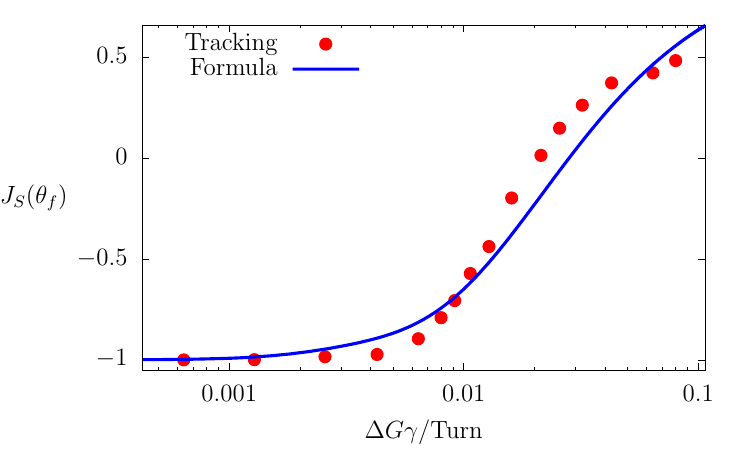}
\caption{\label{fig:RHIC_FS}Comparison of the final spin action $J_S(\theta_f)$ from tracking through the resonance in Fig.~\ref{fig:RHIC_FS_ADST} and the asymptotic spin action predicted by Eq.~\eqref{eq:Two_Slope_FS}.}
\end{figure}

We track through the resonance shown in Fig.~\ref{fig:RHIC_FS_ADST}. We start tracking at $G\gamma(\theta_i)=384.36$ with $J_S(\theta_i)=1$ using an ISF computed by stroboscopic averaging. We stop tracking at $G\gamma(\theta_f)=385$ and compute $J_S(\theta_f)$ by stroboscopic averaging. We again use Eq.~\eqref{eq:Two_Slope_FS} to approximate $J_S(\theta_f)$. The result is shown in Fig.~\ref{fig:RHIC_FS}. There is noticeable disagreement, which is not surprising as the tracking begins at an energy with $P_\mathrm{lim} \approx 45\%$ and ends at an energy with $P_\mathrm{lim} \approx 65\%$. Thus, we are \emph{not} tracking from far before an isolated resonance to far after an isolated resonance. However, Eq.~\eqref{eq:Two_Slope_FS} provides a reasonable estimate of $J_S(\theta_f)$.

For large ramping rates, it appears that $J_S(\theta_f)$ does not approach 1 as predicted by Eq.~\eqref{eq:Two_Slope_FS}, but rather levels off at a smaller value. One possible explanation for this phenomenon is that, in arguing for the applicability of the Froissart-Stora formula and Eq.~\eqref{eq:Two_Slope_FS}, we assumed that the product $\alpha\eta_3\hat{s}$ was small, which may be violated in the case of fast ramping.

\begin{figure}
\centering
\includegraphics[scale=0.65]{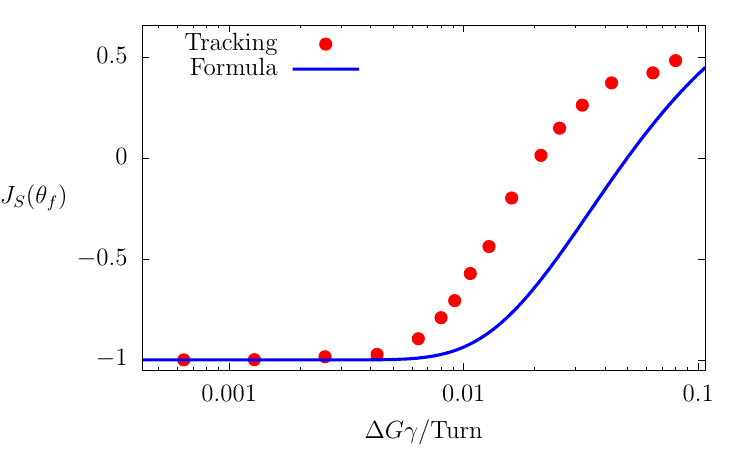}
\includegraphics[scale=0.65]{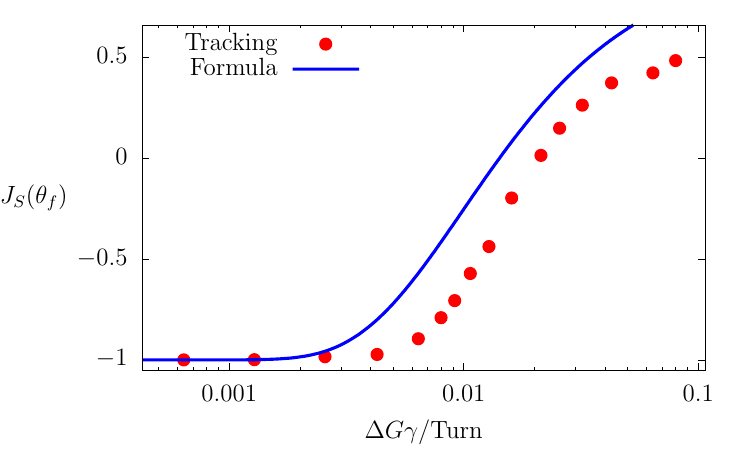}
\caption{\label{fig:RHIC_FS_ONE_SLOPE}Comparison of the final spin action $J_S(\theta_f)$ after tracking through the  resonance in Fig.~\ref{fig:RHIC_FS_ADST} and the asymptotic spin action predicted by the Froissart-Stora formula. The agreement is far inferior to that of Fig.~\ref{fig:RHIC_FS}. Top: Inserting the slope before the resonance crossing into the Froissart-Stora formula. Bottom: Inserting the slope after the resonance crossing into the Froissart-Stora formula.}
\end{figure}

Before the introduction of Eq.~\eqref{eq:Two_Slope_FS}, it was only possible to analytically estimate the polarization loss due to higher-order resonances for which $\dv*{\nu}{\theta}$ was approximately the same before and after the resonance crossing. However, the shape of $\nu(\theta)$ generally has a complicated dependence on the lattice layout so that only a small subset of higher-order resonances satisfy this requirement. Thus, it was not clear how to apply the Froissart-Stora formula to higher-order resonances such as those shown in Figs. \ref{fig:Res_ex} and \ref{fig:RHIC_FS_ADST}, where the slope of $\nu$ changes drastically at the resonance crossing. One possibility was to choose either the slope before the resonance or the slope after the resonance for use in the Froissart-Stora formula. However, as shown in Fig.~\ref{fig:RHIC_FS_ONE_SLOPE}, this method generally leads to very poor estimates of $J_S(\theta_f)$. In fact, the complicated form of Eq.~\eqref{eq:Two_Slope_FS} demonstrates that there is no ``effective slope" which could have been used in the original Froissart-Stora formula.

\section{Summary}
The use of the Froissart-Stora formula to calculate the polarization loss caused by first-order spin-orbit resonance crossings is well-established and in rings with well-separated resonances gives satisfactory results without knowledge of the invariant spin field or the amplitude-dependent spin tune. A naive extension of this theory, based on the closed-orbit spin tune alone, would lead one to believe that there can be no spin-orbit resonance crossings in rings with Siberian snakes as long as the orbital tunes are chosen appropriately. However, the persistence of polarization loss in such rings cannot be ignored. 

Although it has regrettably not been widely recognized, it was shown more than twenty years ago that such polarization loss can be explained by higher-order spin-orbit resonance crossings and that the Froissart-Stora formula can be used together with the invariant spin field and the amplitude-dependent spin tune to predict the corresponding depolarization~\cite{Higher_Order}. However, the usual Froissart-Stora formula suffers from limited applicability in this case because the slope of the amplitude-dependent spin tune often changes at the moment of resonance crossing. This work presents an extension to the Froissart-Stora formula which allows for such a change in slope and demonstrates its applicability. In the example presented, the new formula leads to significantly better agreement with tracking than {that from} simply using one of the slopes in the original Froissart-Stora formula.

The new formula presented herein is still not perfectly accurate for various reasons, including the fact that the amplitude-dependent spin tune is not actually a piecewise-linear function and that higher-order resonances almost always have a nearby neighbor. Future work could investigate extensions to the Froissart-Stora formula which account for these factors.

\begin{acknowledgments}
The authors acknowledge many helpful discussions with E.~Hamwi, K.~Heinemann, J.~A.~Ellison, and H.~S.~Dumas. This work was supported by the U.S. Department of Energy under award number DE-SC0012704. The work of J.~P.~D. was additionally supported by the U.S. Department of Energy, Office of Science, High Energy Physics, under award number DE-SC0024907, the Tigner Traineeships in Accelerator Science program.
\end{acknowledgments}

\bibliography{refs.bib}

\end{document}